\documentclass{emulateapj}
\shorttitle{HRI ULX Statistics}
\shortauthors{Ptak \& Colbert}

\newcommand {\ergs} {erg s$^{-1}$}
\newcommand {\ergcms} {erg cm$^{-2}$ s$^{-1}$}
\slugcomment{Accepted by ApJ}
\begin{document}

\title{The Statistical Properties of Galaxies Containing ULXs}
\author{A. Ptak\altaffilmark{1}, E. Colbert\altaffilmark{1,2}}
\altaffiltext{1}{The Johns Hopkins University, Homewood Campus,
  Baltimore, MD 21218, ptak@pha.jhu.edu}
\altaffiltext{2}{The Catholic University of America, 620 Michigan Ave NE, 
Washington, DC 20064, colbert@cua.edu}

\begin{abstract}
We present a statistical analysis of the properties of galaxies
containing ultraluminous X-ray objects (ULXs).  Our primary goal is to
establish the fraction of galaxies containing a ULX as a function of
ULX luminosity.  
Our sample is based on {\it ROSAT} HRI observations of galaxies.  We
find that 
$\sim 12\%$ of galaxies contain at least 
one ULX with $L_{X} > 10^{39}$ \ergs and $\sim 1\%$ of galaxies contain
at least one ULX with $L_{X} > 10^{40}$ \ergs. These ULX frequencies are
lower limits since ROSAT HRI observation would
miss absorbed ULXs (i.e., with $N_H \ga 10^{21} \rm \ cm^{-2}$) and
those within $\sim 10$'' of the nucleus (due to the positional error
circle of the {\it ROSAT} HRI).  
The Hubble type distribution of galaxies with a ULX differs
significantly from the distribution of types for nearby RC3 galaxies,
but does not differ significantly from the galaxy type distribution of
galaxies observed by the HRI in general.  We find no increase in the
mean FIR luminosity or FIR / K band luminosity ratio for galaxies with
a ULX relative to galaxies observed by the HRI in general, however
this result is also most likely biased by the soft bandpass of the HRI
and the relatively low number of high SFR galaxies observed by the HRI
with enough sensitivity to detect a ULX.
\end{abstract}

\section{Introduction}
{\it ROSAT} HRI observations showed that 
ultraluminous X-ray objects (ULXs, a.k.a.  IXOs), 
which were previously detected with {\it Einstein}
\citep[see][]{fa89}, are compact X-ray sources, are quite common
in the local Universe, and are typically
offset from the optical 
galactic nucleus \citep{co99, ro00}.
Here we define ULXs as X-ray sources in
galaxies that are point-like, extra-nuclear, and have X-ray
luminosities in excess of $10^{39} \ \rm ergs \ s^{-1}$.
%More complete surveys (e.g. Colbert
%\& Ptak 2002) from {\IT ROSAT} and {\it Chandra} place their
%frequency of occurrence at $\sim$1 in every 5 galaxies.  
There is considerable debate as to nature of these sources
\citep[see][for reviews]{va03,mi03}, with popular explanations including
anisotropic emission from stellar-mass X-ray binaries and
intermediate mass black holes. 
 A constraint on any explanation for
these sources would be the frequency at which they occur, i.e., what
  fraction of galaxies harbor a ULX as a function of ULX luminosity.
For example, this frequency would have implications for the typical
relativistic boosting factor for jet models of ULXs \citep{ko02}.  In
this paper we examine the frequency of ULX occurrence as a function of
ULX luminosity based on the HRI ULX catalog published in \citet{co02}.
While HRI observations suffer from significant biases against the
detection of ULXs (as discussed below), we note that the relatively
large field of view of the HRI 
resulted in a large number of galaxies having been observed, and of
course large area surveys are crucial for establishing the bright end
of a luminosity distribution.

\section{Methodology \label{methsec}}
In addition to \citet{co02}, we
also made use of a subset of the RC3 galaxy catalog \citep{de91} where the
distance is determined, most often based on \citet{tu88}.
\citet{co02} chose a recessional velocity cut-off of 5000 km s$^{-1}$.
 In this paper we assume the cosmology $H_0$ = 75 km s$^{-1}$
 Mpc$^{-1}$  and $q_0$ = 0.5 (for consistency
with Tully 1988), and therefore this velocity corresponds to 66.7 Mpc.  
This resulted in a galaxy sample
containing 9452 galaxies.  We next determined which of these galaxies had
``coverage'' by computing the overlap of the RC3 $D_{25}$ ellipse and
the HRI FOV for each observation that was used in \citet{co02}.  We
considered a galaxy as having coverage if $\ge 90$\% of the galaxy was in the
FOV.  We found that 766 galaxies had coverage, or $\sim 8\%$ of the
RC3 subset.

In order to properly assess the frequency of ULXs, we need to assess
how many galaxies were 
observed by the HRI with sufficient sensitivity to detect a ULX.
We computed the flux limit at the position of each RC3 galaxy with
coverage by computing the background level in the point source region
size.  The 
point source region used by  XAssist \citep{pt03} is a circle with a
radius given by $4[\exp(\theta/9) + 1]$ arcseconds, where $\theta$ is the
off-axis angle of the source in arcminutes.  The total number of
background counts, b, in
this area was then determined using the median
background counts per pixel (also computed as a part of the XAssist
processing of each field).   The $3\sigma$
detection threshold count rate is given by $3\sqrt{b}/t$, where t is
the exposure time, and the limiting
flux is then $3\sqrt{b}f_{conv}v(\theta)/t$ where $v(\theta)$ gives the
vignetting correction and $f_{conv}$ is the on-axis count rate to flux
conversion factor.  $f_{conv}$ was computed using the FTOOL PIMMS,
assuming a power-law slope of $1.7$ and the Galactic column density
\citep[using the FTOOL ``nh'']{di90}.   
We computed the limiting flux in both the 0.5-2.0 keV
bandpass $F_{\rm \ lim, 0.5-2.0 \ keV}$ and the 2-10 keV bandpass
$F_{\rm \ lim, 2-10 \ keV}$.  We used 
$F_{\rm \ lim, 2-10 \ keV}$ to compute the limiting luminosity since
the 2-10 keV bandpass was used in \citet{co02}, for consistency with
current {\it Chandra} and {\it XMM-Newton} surveys.  Any ULX could
then have been detected in a galaxy when $L_{\rm \ lim, 2-10 \ keV}$
is $\leq 10^{39}$\ergs, while more generally $L_{\rm \ lim, 2-10 \
  keV}$ is $>10^{39}$\ergs. 

Finally, it is necessary to determine how many ULXs are likely to
be background sources.  For this purpose we used the logN-logS given
in \citet{ha98}, which we note is consistent with a simple power-law
$\log N(>S) \sim 100(S_{-14})^{-1.4}$ sources deg$^{-2}$, where
$S_{-14}$ is the 0.5-2.0 keV flux in units of $10^{-14} \rm \ ergs \
cm^{-2} \ s^{-1}$. The number of expected background sources is then
$n_{xrb} = 100(\frac{F_{\rm \ lim, 0.5-2.0 \ keV}}{10^{-14}})^{-1.4}
  \pi r^2$, where r is the major axis radius of each galaxy in
  degrees.  \citet{co02} used $r=D_{25}$ (the RC3 major axis
  diameter) however here we also consider $r =
  0.5D_{25}$.   Also note that when $L_{\rm \ lim, 2-10 \ keV}$ is
  $\leq 10^{39}$\ergs, the relevant number of background
  sources is the number of source whose {\it apparent} luminosity
  exceeds $10^{39}$\ergs, and since we are assuming a simple power-law
  logN-logS this simply entails decreasing the total number of
  background sources expected by the factor ($L_{\rm \ lim,
    2-10 \ keV}/10^{39}$\ergs)$^{1.4}$.

\section{Results}
%The main empasis of this paper concern the properties of galaxies that
%contain at least one ULX.  In what follows, ``background subtracted''
%galaxy counts refer to the number of galaxies where the numer of ULXs
%exceed the number of background sources.

\subsection{ULX Frequency}
In Table \ref{cntstab} we show the number of galaxies ($N_{L_{lim} \le
  L_X}$) with HRI 
coverage and a limiting 2-10 keV luminosity {\it less than} or equal to
$L_X$ ranging from $10^{39}$\ergs to $10^{41}$\ergs (i.e., the number
of galaxies where a ULX as bright as $L_X$ or brighter could have been
detected).  We also show the number of galaxies with at least
one ULX ($N_{wULX}$), and the number of
galaxies with at least one ULX whose luminosity exceeds the 
limiting luminosity ($N_{L_{ULX} \ge L_X}$).  This latter quantity,
$N_{L_{ULX} \ge L_X}$, is of interest since it represents the number
  of galaxies 
with a ULX at least as bright as the list $L_X$, and 
we then computed the resultant number of galaxies after subtracting the
expected number of background sources ($N_{net, L_{ULX} \ge L_X}$, i.e., the
number of galaxies where the 1-$\sigma$ lower limit on the number of
  ULXs is greater than 0.; the lower limit was calculated using the
  mehology described in Kraft, Burrows, \& Nousek (1991), with the
  background being the 
  expected number of background sources).  Finally
we list the values $N_{wULX}$, $N_{L_{ULX} \ge L_X}$, and $N_{net,
  L_{ULX} \ge L_X}$ 
using both $r=D_{25}$ and $r=0.5D_{25}$.  Note that in the latter case
the total number of ULXs in the \citet{co02} sample is reduced from 87
to 49.

\begin{figure}
%\plotone{ulx_galaxy_cnts_27oct03.eps}
\plotone{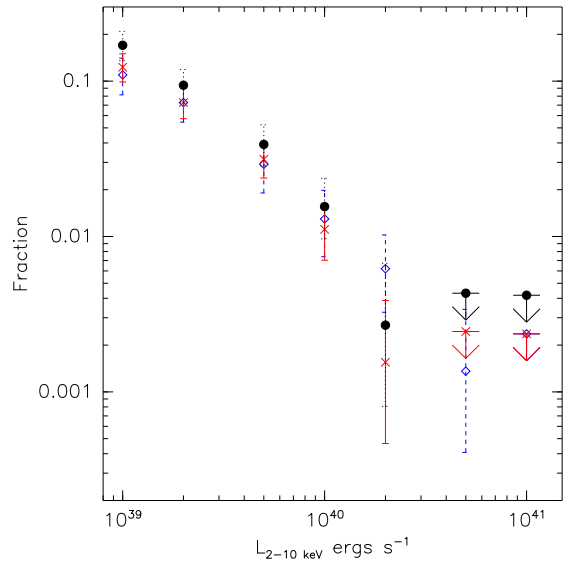}
\caption{The frequency of ULX occurrence in galaxies as a function of
  ULX luminosity. The frequency of ULXs is plotted using effective
  galactic radii of $D_{25}$ (blue diamonds with dashed error bars) and
  0.5$D_{25}$ (red crosses with solid error bars). We also plot the
  frequency observed after limiting the sample to only spirals (with
  Hubble type $>=$ 0 and radius = 0.5$D_{25}$; black filled circles with
  dotted error bars). 
\label{fracfig}}
\end{figure}

In Table \ref{cntstab} shows that a large fraction of ULXs detected
at radii larger $0.5D_{25}$ are consistent with background sources,
i.e., the {\it net} number of galaxies with ULXs is often larger when
$0.5D_{25}$ is chosen as the limiting radius.  Also, in some cases,
the number of ULXs detected within the $0.5D_{25}$ radius is
above the background level when $r=0.5D_{25}$ but is
not above the background level when $r=D_{25}$ (i.e., since in the
latter case the expected number of background sources is a factor of
four larger). 
In Figure \ref{fracfig} we plot the fraction of galaxies harboring a
ULX as a function of ULX luminosity 
($N_{net, L_{ULX} \ge L_X}/N_{L_{lim} \le L_X}$) for
both $r=D_{25}$ and $r=0.5D_{25}$.  Here the errors were computed according to
the methodology described in \citet{kr91}, with the background taken as
the number of galaxies at each luminosity that are consistent with all
ULXs being background sources (i.e., 
$N_{L_{ULX} \ge L_X} - N_{net, L_{ULX} \ge L_X}$).
The choice of effective radius for the galaxies does not significantly
impact the estimate of ULX frequency once background sources are taken
into account.  The frequency of ULXs observed by the HRI evidently
is $\sim 12\%$ and $\sim 1\%$ of
galaxies harboring a ULX with $L_X >= 10^{39}$\ergs and $L_X >=
10^{40}$\ergs, respectively.  We also plot the galaxy fractions for
spirals only (i.e., Hubble type in RC3 $\ge$ 0).  The Hubble type
distribution of galaxies with and without a ULX is discussed below,
however we note that the fraction of galaxies harboring a ULX is
slightly higher in spirals.

%\clearpage

\begin{deluxetable*}{ll|lll|lll}
\tabletypesize{\scriptsize}
\tablecaption{Galaxy Number Counts\label{cntstab}}
\tablehead{
& & \multicolumn{3}{c}{$r=D_{25}$} &
\multicolumn{3}{c}{$r=0.5D_{25}$}\\
\colhead{$L_X$} & \colhead{$N_{L_{lim} \le L_X}$} 
& \colhead{$N_{wULX}$} &
\colhead{$N_{L_{ULX} \ge L_X}$} & 
\colhead{$N_{net, L_{ULX} \ge L_X}$}
& \colhead{$N_{wULX}$} &
\colhead{$N_{L_{ULX} \ge L_X}$} & 
\colhead{$N_{net, L_{ULX} \ge L_X}$}\\
\colhead{(1)} & \colhead{(2)} & \colhead{(3)} & \colhead{(4)} &
\colhead{(5)} & \colhead{(6)} & \colhead{(7)} & \colhead{(8)}
}

\startdata
% Revised 1/16/04
%1 &  228 & 45 & 45 & 27 (11.8\%) & 34 & 34 & 31 (13.6\%)\\
%2 &  316 & 49 & 38 & 19 (6.0\%) & 37 & 27 & 24 (7.6\%)\\
%5 & 446 & 53 & 23 & 8 (1.8\%) & 38 & 14 & 13 (2.9\%)\\
%10 & 540 & 54 & 11 & 5 (0.9\%)& 38 & 6 & 5 (0.9\%)\\
%50 & 736 & 54 & 1 & 1 (0.1\%)& 38 & 0 & 0 (0.0\%)\\
%100 & 762 & 54 & 0 & 0 (0.0\%)& 38 & 0 & 0 (0.0\%)\\
1 &  228 & 45 & 45 & 25 (8.7\%) & 34 & 34 & 28 (12.3\%)\\
2 &  316 & 49 & 38 & 23 (7.3\%) & 37 & 27 & 23 (7.3\%)\\
5 & 446 & 53 & 23 & 13 (2.9\%) & 38 & 14 & 14 (3.1\%)\\
10 & 540 & 54 & 11 & 7 (1.3\%) & 38 & 6 & 6 (1.1\%)\\
50 & 736 & 54 & 1 & 1 (0.1\%) & 38 & 0 & 0 (0.0\%)\\
100 & 762 & 54 & 0 & 0 (0.0\%) & 38 & 0 & 0 (0.0\%)\\
\enddata
\tablecomments{(1) minimum ULX 2-10 keV luminosity in $10^{39}$\ergs. (2)
  $N_{L_{lim} \le L_X}$ = number of galaxies with a limiting
  luminosity $L_{lim}$ less than or equal to $L_X$. (3) and (6)
  $N_{wULX}$ = number of galaxies with $L_{lim} \le L_X$ and a ULX,
  for $r=D_{25}$ and $r_{0.5D_{25}}$, respectively.
  (4) and (7) $N_{L_{ULX} \ge L_X}$ = number of galaxies with a
  $L_{lim} \le L_X$ and a ULX with luminosity $L_{ULX} \ge L_{X}$.
  (5) and (8) $N_{net, L_{ULX} \ge L_X}$ = number of galaxies with a
  $L_{lim} \le L_X$ and a net number of ULXs with luminosity
  $L_{ULX} \ge L_{X}$ $>$ 0 after subtracting the expected number of
  background sources, with the percentage of galaxies (i.e., $N_{net,
    L_{ULX} \ge L_X} / N_{L_{lim} \le L_X}$) given in parentheses.}
\end{deluxetable*}

%\clearpage

%\clearpage

\subsection{Hubble Type Distribution}
\begin{figure*}
%\plottwo{hri_galtype_histogram_norm_d25_08oct03.eps}{hri_galtype_hri_ulx_werrors_d25_24oct03.eps}
\plottwo{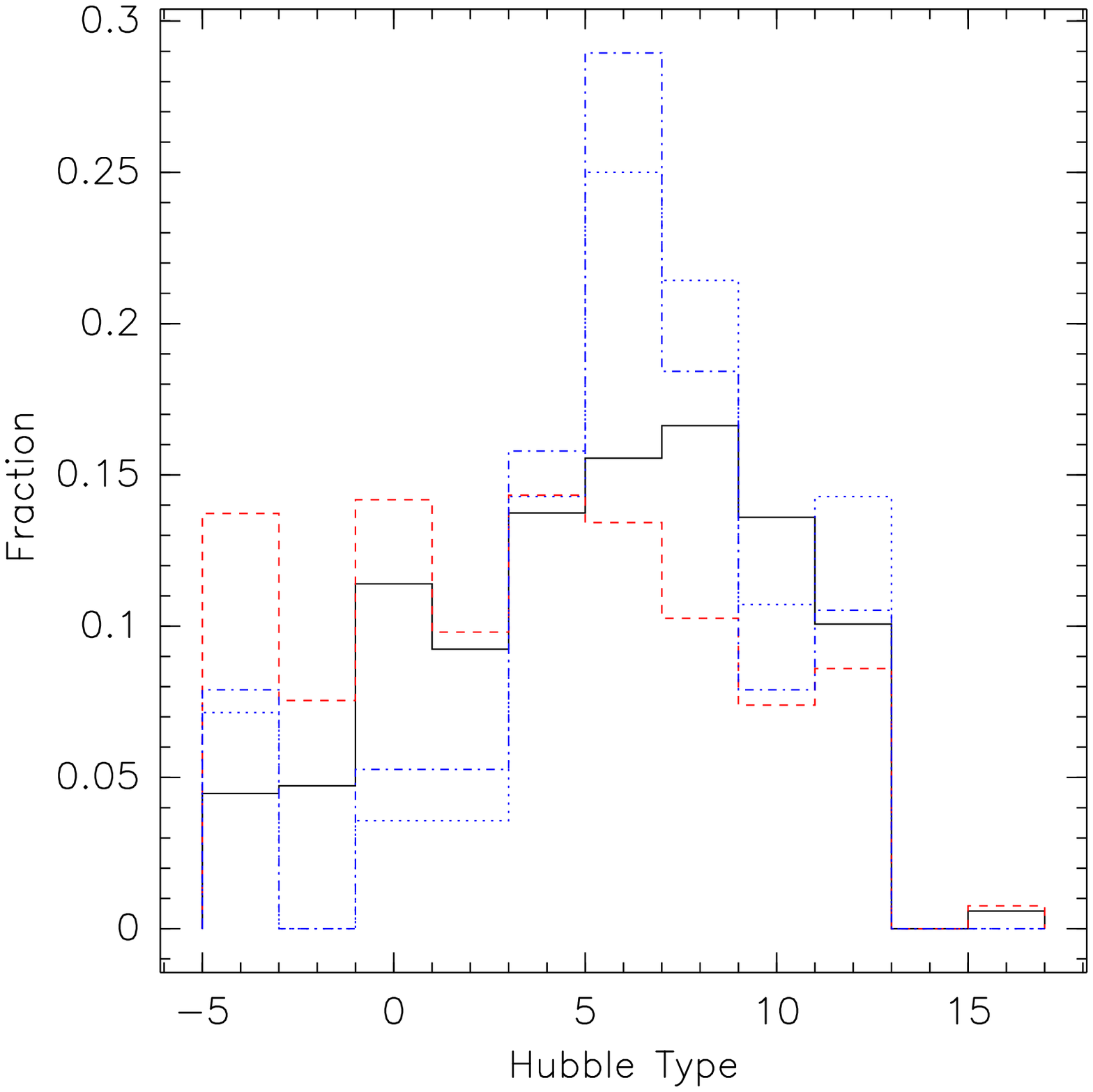}{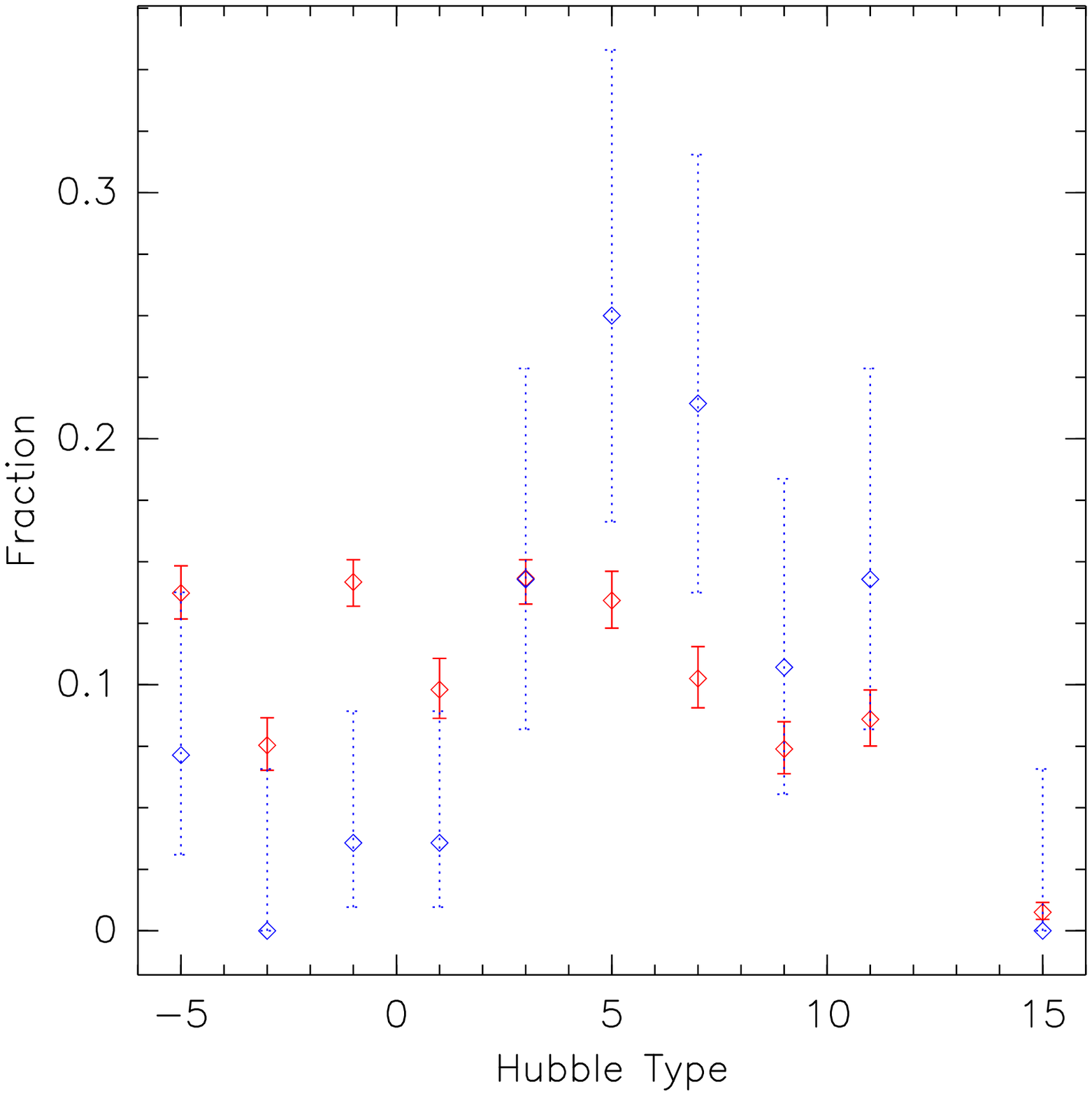}
\caption{(left) The Hubble type distribution of nearby galaxies in the RC3 catalog
(black solid line), the galaxies with HRI coverage (red
dashed line), the galaxies with a ULX (blue dot-dashed line) and the
galaxies with a ULX after excluding galaxies with at least one ``net'' ULX
(i.e., after subtracting the expected number of background sources;
blue dotted line).  (right) The Hubble type histograms for galaxies
with HRI coverage (red, solid error bars) and with a ``net'' ULX (blue,
 dotted error bars) shown with errors computed using \citet{kr91}.
The galaxy types T correspond to: ellipticals have T
=-5 to T=-4, lenticulars have T=-3 to T=-1, spirals have T=0 (S0) to
T=9, with odd numbers from 1 to 9 representing Sa,b,c,d and even
numbers representing Sab,bc,cd,dm.  Irregulars have T=10-11 and we
assigned 16 to galaxies with T$>$11. Both plots were created using only
ULXs within $r=0.5D_{25}$.\label{htypefig}}
\end{figure*}

In Figure \ref{htypefig} we show the Hubble type distribution of the
RC3 subsample (discussed in \S 
%\ref{methsec}
2
), the
Hubble type distribution of galaxies that had HRI coverage (hereafter
HRI galaxies) and the HRI galaxies that contain at least one ULX.  The
ULX galaxy type distribution is shown both
before and after background subtraction, and here we only consider
$r<0.5D_{25}$ ULXs.   There is a slight excess
of elliptical ULX galaxies relative to the RC3 sample however the
larger excess in the HRI galaxy sample suggests that this bias due in
part to a bias toward early type galaxies in the HRI observing
patterns relative to RC3 galaxies in general.  The most conspicuous
feature of the ULX galaxy distribution 
is the excess of T=5-7 galaxies, corresponding to Sc and Scd
galaxies. However, in general the ULX galaxy type distributions are not
significantly different statistically from the HRI RC3 galaxy type
distributions, with the $\chi^2$ tests giving $\chi^2/dof \sim 15/8$,
which can only be rejected at the 95\% level.  

%\clearpage
 
%\clearpage

\begin{figure*}
%\plottwo{hri_gals_fir_sfr_d25_27oct03.eps}{spirals_hri_gals_fir_sfr_d25_27oct03.eps}
\plottwo{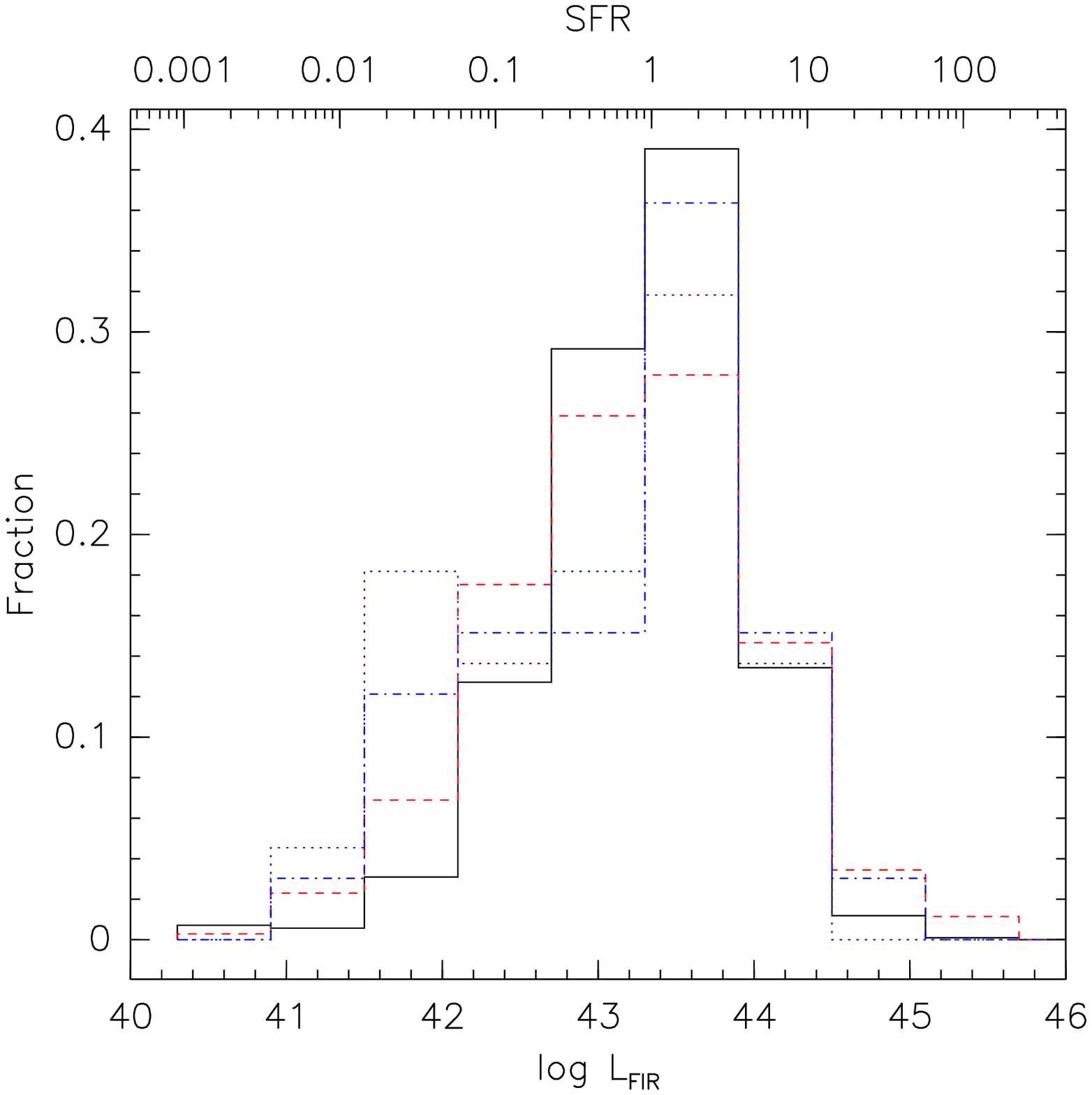}{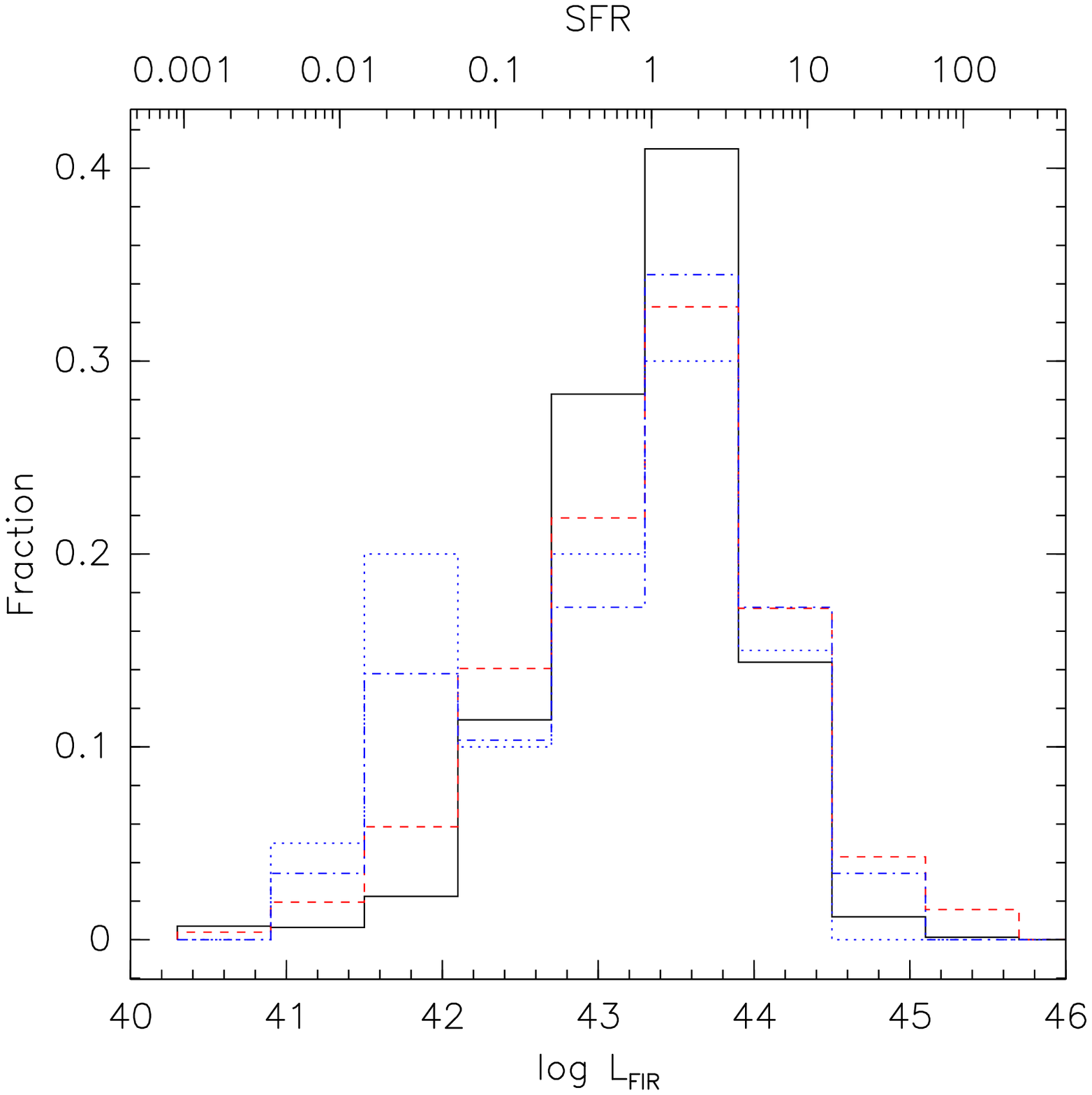}
\caption{The FIR luminosity distribution of nearby galaxies with IRAS
  detections at 60 $\mu m$ and 100 $\mu m$ (black solid line), the galaxies
  with HRI coverage (red dashed line), and the galaxies with a ULX before
  (blue dot-dashed line) and after (blue dotted line) background
  subtraction. The left panel shows the results without selecting on
  galaxy type and the right panel shows the results when selecting
  only spiral galaxies (Hubble type $\ge$ 0). \label{firfig}}
\end{figure*}

We also investigated the possible connection between star-formation
rate (SFR) and ULX frequency by tabulating the far-infrared luminosities (a
well-known SFR indicator; Kennicutt 1998) of
the galaxy samples discussed above.  These distributions are shown in 
Figure \ref{firfig}, where we used NED IRAS fluxes densities at 60
and 100 $\mu m$ to compute $F_{FIR} = 1.26 \times 10^{-23}[2.58 \times
  10^{-12}f_{60 \mu m} + 10^{-12}f_{100 \mu m}]$.  The peak of the
  distributions with 
a ULX is slightly skewed toward higher FIR (i.e., near $\log L_{FIR}
= 43.5$) however the means of the distributions do not differ  
  significantly (all are in the range log $L_{FIR}$ = 42.7-43.0).  The
  HRI galaxy FIR luminosity distribution differs from the ``net'' ULX
  galaxy FIR luminosity distribution at less than the 90\% level based
  on the $\chi^2$ test.
Restricting the sample to spiral galaxies (T $>=$ 0) shifted the means
by only $\log
L_{FIR} \sim 0.1$, reflecting the predominance of spirals in these samples.

%\clearpage

%\clearpage

\begin{figure*}
%\plottwo{hri_gals_fir_fk_d25_27oct03.eps}{spirals_hri_gals_fir_fk_d25_27oct03.eps}
\plottwo{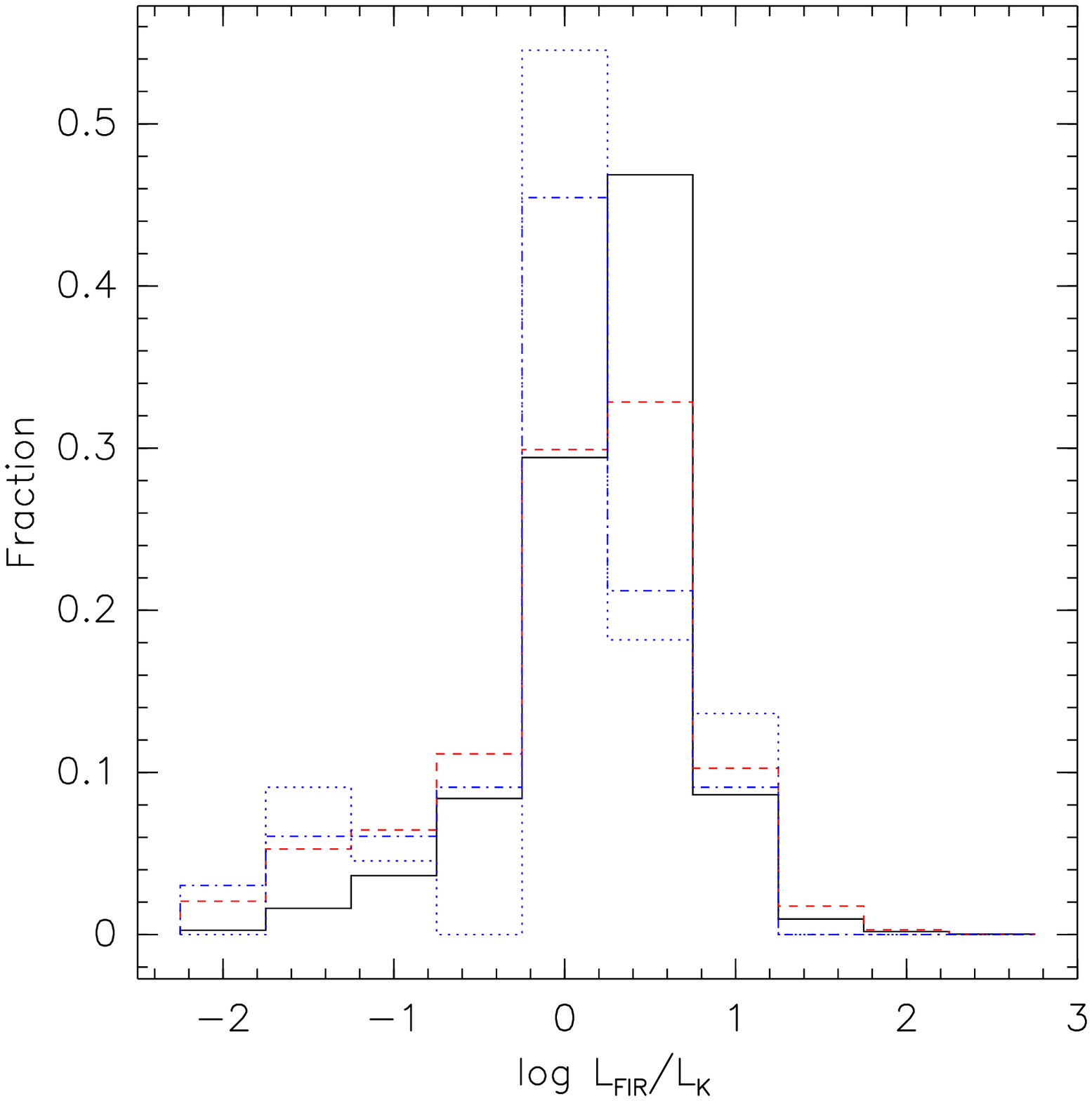}{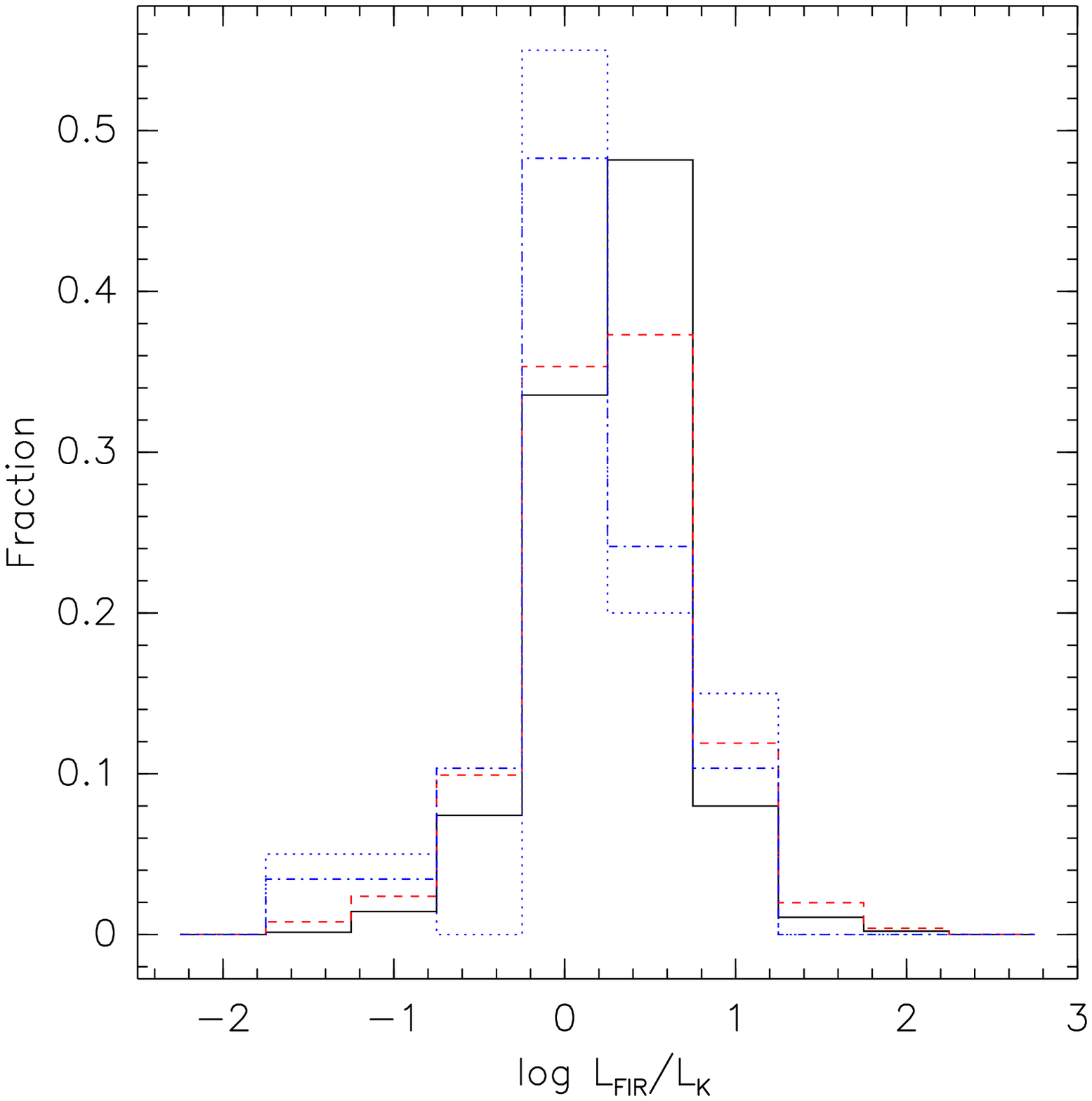}
\caption{The FIR/K band luminosity distributions for nearby galaxies with IRAS
  detections at 60 $\mu m$ and 100 $\mu m$ and 2mass extended source
  catalog K band detections (black solid line).  Other curves are as in
  Figure \ref{firfig}. The left panel shows the results without selecting on
  galaxy type and the right panel shows the results when selecting
  only spiral galaxies (Hubble type $\ge$ 0). \label{firfkfig}}
\end{figure*}

Similarly we plot the distribution of $L_{FIR} / L_{K}$ where $L_K$ is
the K band luminosity computed using the 2mass Extended Source catalog
K band magnitude \citep[see][]{ja03}
and assuming a zero-point flux density of $6.7
\times 10^{-21}$\ergcms$Hz^{-1}$ \citep{zo90}\footnote{see also\\ 
http://www.ipac.caltech.edu/2mass/releases/allsky/doc/sec6\_4a.html}.
The K band luminosity is a proxy for stellar mass since the variation
in mass-to-light ratio for K band is only a factor of $\sim 2$ for
various galaxy types \citep{be01}.  
%This ratio therefore serves to
%remove biases due to larger FIR luminosities due predominantly to larger
%galaxy sizes.
These distributions are shown in Figure \ref{firfkfig} and again we
do not see any significant differences between galaxies with ULXs and
galaxies observed by the HRI in general.

%\clearpage

%\clearpage

\section{Summary \& Conclusions}
We have assessed the frequency of ULX occurrence in galaxies as a
function of ULX luminosity, including proper assessment of the
contamination from background sources.  We have found that assuming an
effective radius of $D_{25}$ tends to result in a significant
contamination from background sources.  We found frequencies of $\sim
12\%$ and  $\sim 1\%$ for ULX luminosities of at least $10^{39}$\ergs and
$10^{40}$\ergs. The Hubble type distribution of galaxies containing at
least one ULX is significantly different from those that do not
contain a ULX in general, however it is not significantly different
from galaxies observed by the HRI.  This implies that any bias in the
Hubble type distribution of galaxies observed by HRI with ULXs is
mostly due to the observing patterns of the HRI.  Recently
\citet{ir04} reported on an analysis of ULXs in early-type galaxies
and found that ``very'' luminous ULXs (with $L_X > 2
\times 10^{39}$\ergs) are consistent with the expected number of
background sources.  This result is based in part on {\it Chandra} data
and uses luminosities in the 0.3-10.0 keV bandpass as opposed to our
choice of the 2-10 keV band.  However, as can be seen in Figure
\ref{htypefig}b, even 
when considering all ULXs in our sample (i.e., with $L_X \ge 1 \times
10^{39}$\ergs),  the frequency of early-type galaxies containing ULXs
is only marginally in excess of zero (with ULXs occurring in
globular clusters as in the case of NGC 1399 contributing to the
early-type galaxy ULX frequency). 

The main caveat with an analysis of the HRI observations is the
softness of its bandpass (i.e., 0.1-2.2 keV).  Objects absorbed by
column densities in excess of $\sim 10^{21} \ \rm cm^{-2}$ are
therefore difficult with the HRI.  Since columns of this order are
typically observed through the disks of galaxies, it is likely that
ULXs are being missed that are close to the nucleus of the galaxies
and/or are observed in edge-on disks.  {\it Chandra} and {\it XMM-Newton}
observations will therefore be much more adept at detecting ULXs in
these types of environments.  This is particularly true for {\it Chandra}
whose much smaller PSF relative to the HRI and {\it XMM-Newton} allows
nuclear ULX to be 
discerned from AGN at smaller offsets (i.e., 1'' compared with 10'').
\citet{si03} finds that 8/41 $\sim 20$\% of nearby galaxies in a
relatively unbiased {\it Chandra} survey harbor a
ULX.  This suggests that our HRI survey is missing of order $\sim
40\%$ of ULXs (albeit with large uncertainty due to the size of the
Chandra sample).  \cite{ho03} reports that the frequency of ULXs with
$L_X >= 2 \times 10^{39}$\ergs 
is $\sim 8\%$ locally and $\sim 36\%$ at z $\sim 0.1$ \citep{ho03}
(suggesting that the frequency of ULXs may evolve).    
We are similarly finding $\sim 8\%$ (see Table \ref{cntstab}) which
implies that any biases in our survey are not particularly large
for $L_X >= 2 \times 10^{39}$\ergs ULXs, within the errors.  

The lack of any obvious connection between FIR luminosity (and hence
SFR) or the FIR/K band luminosity ratio distribution and 
the presence of a ULX suggests that if such a
connection exists, then the biases discussed above are particularly acute in
star-forming galaxies.  This may not be surprising given that
high-SFR galaxies tend to contain significant amounts of molecular
clouds and dust which would obscure soft X-rays.  In addition the
relatively small size of our sample may also be impacting this
result.  For example, only 23 galaxies have $\log L_{FIR} > $43.5
\ergs and enough HRI sensitivity to detect a ULX, and 4 harbor a
ULX (after ``background subtraction'' and 
using a galaxy radius of $0.5D_{25}$).  This amounts to a larger ULX
frequency of $17^{+12}_{-8}$\%, although still statistically
consistent with the full sample result.  
Recent {\it Chandra}
surveys of the point source luminosity functions in galaxies suggest
that, particularly at high star-formation rates, the point source
luminosity function normalization is correlated with the SFR (Grimm et
al. 2003, Colbert et al. 2004), and that the luminosity function is
flatter in spirals \citep{co04} or (similarly) in higher SFR galaxies
\citep{ki02}.  Having both a
flatter luminosity function slope and larger normalization at higher
SFR strongly implies that higher SFR galaxies (i.e., with SFR greater
than several to ten $M_{\odot} \ \rm yr^{-1}$ based on Colbert et
al. 2004) would tend to harbor {\it multiple} ULXs.  
This is also suggested anecdotally by the large numbers of ULXs found
in merging galaxies such as the Antennae \citep{fa01}.  Accordingly the
distribution of $L_{FIR}$ and $L_{FIR}/L_K$ for galaxies with {\it at
  least one} ULX may not be significantly impacted by an increase in
the total number of ULXs in galaxies with large SFR.
However the resolution of these issues will ultimately require more
detailed statistical analysis based on {\it Chandra} and {\it
  XMM-Newton} wide-area surveys.   

\acknowledgements
This work was supported by NASA grant NAG 5-11670.  We made use of the
High Energy Archive and Research Center at NASA/GSFC and the NASA
Extragalactic Database.  We thanks the anonymous referee for useful
comments.


\begin{thebibliography}{}
\bibitem[Bell \& de Jong(2001)]{be01}Bell, E. \& de Jong, R. 2001,
  \apj, 550, 212
\bibitem[Colbert et al.(2004)]{co04}Colbert, E., Heckman, T., Ptak,
  A., Strickland, D. \& Weaver, K. 2004, \apj, in press, astro-ph/0305476
\bibitem[Colbert \& Ptak(2002)]{co02}Colbert, E. \& Ptak, A. 2002,
  \apjs, 143, 25
\bibitem[Colbert \& Mushotzky(1999)]{co99}Colbert, E. \& Mushtozky,
  R. 1999, \apj, 519, 89
\bibitem[de Vaucouleurs et al.(1991)]{de91} de Vaucouleurs, G., de
  Vaucouleurs, A., Corwin, H. G., Jr., Buta, R. J., Paturel, G., \&
  Fouqué, P. 1991, Third Reference Catalog of Bright Galaxies (New
  York: Springer) (RC3)
\bibitem[Dickey \& Lockman(1990)]{di90}Dickey, J. \& Lockman, F. 1990,
    Ann. Rev. Ast. Astr. 28, 215
\bibitem[Fabbiano(1989)]{fa89}Fabbiano, G. 1989, \araa, 27, 87
\bibitem[Fabbiano, Zezas, \& Murray(2001)]{fa01}Fabbiano, G., Zezas,
  A. \& Murray, S. 2001, \apj, 554, 1035
\bibitem[Gehrels(1986)]{ge86}Gehrels, N. 1986, \apj, 303, 336
\bibitem[Grimm et al.(2003)]{gr03}Grimm, H., Gilfanov, M., Sunyaev, R. 2003,
  \mnras, 339, 793
\bibitem[Hasinger(1998)]{ha98}Hasinger, G. 1998, AN, 319, 37
\bibitem[Hornschemeier et al.(2003)]{ho03}Hornschemeier, A. et
  al. 2003, \apjl, in press, astro-ph/0308408
\bibitem[Irwin et al. (2004)]{ir04}Irwin, J., Bregman, J., \& Athey,
  A. 2004, \apjl,  in press, astro-ph/0312393)
\bibitem[Jarrett et al.(2003)]{ja03}
Jarrett, T., Chester, T., Cutri, R., Schneider, S. E.,\& Huchra,
J. 2003, \aj, 125, 525
\bibitem[Kennicutt(1998)]{ke98}Kennicutt, R. 1998, \araa, 36, 189
\bibitem[Kilgard et al.(2002)]{ki02}Kilgard, R. et al. 2002, \apj,
  573, 138
\bibitem[Koerding, Falcke, \& Markoff(2002)]{ko02}Koerding, E.,
  Falcke, H., \& Markoff, S. 2002, \aap, 382, 13
\bibitem[Kraft, Burrows, \& Nousek(1991)]{kr91}Kraft, R., Burrows, D.,
  \& Nousek, J. 1991, \apj, 374, 344
\bibitem[Miller \& Colbert(2003)]{mi03}Miller, M. C. \& Colbert,
  E. 2003, International Journal of Modern Physics D, submitted,
  astro-ph/0308402 
\bibitem[Ptak \& Griffiths(2003)]{pt03}Ptak, A. \& Griffiths, R. 2003, 
  Astronomical Data Analysis Software and Systems XII ASP Conference
  Series, H. E. Payne, R. I. Jedrzejewski, and R. N. Hook, eds., 295,
  465 
\bibitem[Roberts \& Warwick(2000)]{ro00}Roberts, T. \& Warwick,
  R. 2000,\mnras, 304, 52
\bibitem[Sipior(2003)]{si03}Sipior, M. 2003, Ph.D. Thesis, Pennsylvania
  State University
\bibitem[Tully(1988)]{tu88}Tully, R. B. 1988, Nearby Galaxies Catalog
  (Cambridge: Cambridge Univ. Press)
\bibitem[van der Marel(2003)]{va03} van der Marel, R. 2003, Carnegie
  Observatories Astrophysics Series, Vol. 1: Coevolution of Black
  Holes and Galaxies, ed. L. C. Ho (Cambridge: Cambridge Univ. Press),
  in press, astro-ph/0302101
\bibitem[Zombeck(1990)]{zo90}Zombeck, M. V. 1990, Handbook of
  Astronomy and Astrophysics, Second Edition (Cambridge, UK: Cambridge
  University Press)
\end{thebibliography}
\end{document}